# pH-dependent coarse-grained model of peptides

Marta Enciso, Christof Schütte, and Luigi Delle Site

*Institute for Mathematics, Freie Universität Berlin, Germany*

We propose the first, to our knowledge, coarse-grained modeling strategy for peptides where the effect of changes of the pH can be efficiently described. The idea is based on modeling the effects of the pH value on the main driving interactions. We use reference data from atomistic simulations and experimental databases and transfer its main physical features to the coarse-grained resolution according to the principle of *"consistency across the scales"*. The coarse-grained model is refined by finding a set of parameters that, when applied to peptides with different sequences and experimental properties, reproduces the experimental and atomistic data of reference. We use the such parameterized model for performing several numerical tests to check its transferability to other systems and to prove the universality of the related modeling strategy. We have tried systems with rather different response to pH variations, showing a highly satisfactory performance of the model.

## I. INTRODUCTION

The study of biological systems through theoretical and computational means requires a subtle balance between describing the complexity of the real world and the desirable simplicity of a satisfactory theoretical model. This view is particularly useful in the case of proteins, where different interactions, hydrophobic, electrostatic, hydrogen bonds, to name a few, are interconnected in a rather non trivial way. It is widely accepted that some of these interactions drive a particular aspect of the protein overall behavior (e.g. hydrophobic interactions are mainly responsible of tertiary structure formation, hydrogen bonds stabilize secondary structures and aggregates, etc.)[1]. With this in mind, a general picture can be achieved by considering each interaction and its related effect separately and then combine all of them together in a consistent way[2].

An important problem where this line of thought applies is that of the influence of the external conditions (e.g. temperature, concentration, pH, ionic strength) on folding and aggregation[3]. Among the many possible external factors, the role of pH has received a rather extended attention in the last years; in fact it is well known that non-physiological values promote the formation of aggregates[4,5], however its microscopic origin still requires a clear explanation. In this context, the theoretical and computational modeling of the effect of the pH constitutes a major challenge since biomolecules present multiple sites that can release or bind protons and may also be coupled to one another[6].

For this reason, the development of constant-pH simulation methodologies has received a large attention in the last years. A wide variety of methods have been proposed, usually differing in the level of details considered in the model (explicit or implicit solvent, quantum or classical full-atom descriptions) and in describing the proton (continuous or discrete models)[6,7].

Our long term goal is to study folding and aggregation processes for peptides and proteins characterized by long timescales processes, as a function of external factors. By using a modeling strategy based on the consistency across different scales or molecular resolutions. The essential underlying principles of such a strategy consist of extracting information at high (atomistic) level of resolution and then transfer the essential physical features in a proper effective way into the simplified (coarse-grained) model. Next a series of tests of consistency are done by comparing coarse-grained simulations of systems with markable different characteristics from the reference ones with high resolutions simulations or experiments. Finally the coarse-grained model is accepted if the consistency is satisfactory (for example in terms of structures, in case of peptides) or is refined on the basis of this later comparison and a further consistency test is then performed. The procedure is re-iterated until a satisfactory consistency is reached. This sort of strategy has been shown to be rather efficient for a large variety of condensed matter systems[8-15]. In this way, the final simplified model would allow for the sampling of the large scale conformational space, which is relevant for macroscopic properties of these systems, otherwise not accessible (or at highly demanding computational costs) by atomistic simulations. Our coarse-grained protein descriptions is based on a pre-existing and well tested topological description of the peptides with one or few centers of interaction per amino acid[16] on the top of which we then include all the main features related to the changeable external parameter as the pH.

Simulating the change of pH conditions with coarse-grained resolution is an appealing target in the field that so far has not been addressed in a systematic way. Some approaches do use coarse-grained simulations but only as reference potential for free-energy pH-dependent calculations[17]. Others can just simulate extreme pH conditions but not intermediate values[18], being far from the sophisticated full-atom resolution simulations at constant pH[6,7,19]. Given the above described state of the art, our work can be considered in this context a pioneeristic attempt to model and simulate the effects of the change of pH on large systems over extended time scales.

Our starting point in this work is a physics-based model that includes the two main driving forces of peptide systems: hydrogen bonds and hydrophobic interactions[20]. Considering only these two interactions, effects such as temperature, concentration, and, to some extent, sequence can be described; however, it is not sensitive to pH changes. The inclusion of pH-dependent effects instead involves the design of an electrostatic interaction potential that describes the impact of protonation and deprotonation. Then, our aim is to enhance the former model with an additional electrostatic contribution. This coarse-grained model is built such that it is able to give back a realistic response of peptides towards environmental conditions such as pH changes. In addition, it requires the implementation of a constant-pH algorithm to carry out the simulations.

The aforementioned technical aspects of this work are described in Section II. We present the system description and sampling methodology for full-atom and coarse-grained simulations. Then, we evaluate the influence of pH in full-atom simulations, extracting information on the essential features of a pH-dependent potential. After that, we describe the coarse-grained pH-dependent algorithm and the interaction potential that we have used. Section III is devoted to the results of this study. We show three peptides whose structure is pH-dependent and compare our findings with experimental and full-atomistic simulation data, showing a reasonable agreement. Finally, some concluding remarks and future work prospects can be found in Section IV.

## II. METHODS

### A. System description

We describe the peptide system by an off-lattice bead and stick representation with a single center of interaction per amino acid (placed at the $\alpha$-carbon position), with implicit solvent (see Figure 1a). Consecutive beads in the same chain are placed at a constant distance of 0.38 nm (corresponding to a *trans* peptide bond). This level of resolution is widely used by the biophysical community because of its simplicity (and, consequently, its low computational cost)[2]. Although this description may seem rough, it has been proved that the model accurately reproduces the geometry of a natural protein backbone[21].

### B. Sampling method

Atomistic detailed information from peptide systems has been obtained using the GROMACS software[22] and the OPLS/AA force field[23]. For each peptide, we have placed its initial conformation (taken for the PDB database[24]) in a suitable dodecahedron box, where the minimal distance from the peptide to the box wall was 1.5 nm. Next, the molecule was solvated using the SPC/E water model[25] and counterions were added to the system to provide an overall neutral system. With the described set up, energy minimization was performed in order to remove unfavorable contacts and the equilibration was carried out for 500 ps with positional restraints on the peptide atoms; finally, several (1-5) independent 0.5 $\mu s$ simulations for each system were carried out.

Regarding coarse-grained simulations, we have employed a Parallel Tempering Monte Carlo in-house software, parallelized with OpenMP for higher performance. This method is particularly suitable to obtain well-equilibrated ensemble configurations, especially in combination with non-differentiable potentials like ours. We have carried out single-chain and multi-chain numerical experiments using periodic boundary conditions, where each of our simulations presents 24–40 temperatures. Our simulations start from a completely extended conformation for each chain and consists of $8 \cdot 10^6$ Monte Carlo cycles at every temperature after $3 \cdot 10^6$ equilibration cycles (enough to guarantee convergence). In each cycle, every bead of the system is subjected to a trial Monte Carlo move. In order to sample the conformational space as efficiently as possible, we have implemented three individual-bead movements (spike, rotation or translation); besides, each chain is allowed to rotate or translate; in extensive multichain simulations, groups of chains can also rotate or translate together to favor diffusivity of multichain structures (e.g. aggregates). For each system, three or five independent runs have been carried out.

### C. The pH-independent model

In our basic coarse-grained model, the interactions considered correspond to the two main driving forces in peptides: hydrogen bonds and hydrophobic interactions. This model has been described in full detail somewhere else[20,21]. Here we only describe its main features.

The hydrogen bond interaction, $E^{hb}$ is applied between any pair of residues $i$ and $j$ (where $j > i + 2$ and $j \neq i + 4$). If $j = i + 2$, we called it a *local* hydrogen bond; if $j > i + 4$, it is a *non-local* hydrogen bond. Its energy calculation consists of two steps. First, we check three geometrical restrictions (namely, the length of the tentative hydrogen bond between beads, R1; the orientation between the auxiliary vectors, R2; and the relative orientation between those auxiliary vectors and the tentative hydrogen bond, R3), represented in Figure 2. Second, a step-like potential applies if the values of the former restrictions fall within certain limits, expressed in Table I. Acceptable ranges and potential strength differ depending on the kind of hydrogen bond (either local or non-local). The exact definition of the geometrical restrictions as well as the acceptable ranges can be found in Ref.[21].

The hydrophobic interaction, $E^{hg}$, consists in a 10-12 Lennard-Jones potential that distinguishes between hydrophobic, neutral and polar centers of interaction[20,26]. The general expression for an individual interaction follows: $E_{i,j}^{hp} = S_1 \left[ \left( \frac{\sigma_{int}}{r_{i,j}} \right)^{12} - S_2 \left( \frac{\sigma_{int}}{r_{i,j}} \right)^{10} \right]$, where $r_{i,j}$ is the distance between residue $i$ and $j$ and $S_1$, $S_2$ and $\sigma_{int}$ are optimized parameters, shown in Table II.

The chain flexibility is also a key aspect in coarse-grained model design as it determines the accessible conformational space of a given peptide. Some aspects are determined by the peptide definition itself (i.e. the peptide bond that connects amino acids and the size of the amino acids themselves do not allow certain angular combinations). Our algorithm addresses this aspects as part of the Monte Carlo algorithm. In this way, we apply a hard-sphere potential to avoid residue interpenetration. Similarly, we constrain the maximum virtual bond angle between three consecutive residues to a maximum limit of 150° to avoid too extended unnatural conformations.

Other flexibility aspects (for example that related to the question of which conformations are favored in certain conditions) depend very much on the particular definition of the system interactions (particularly, on the way hydrophobic ones are described) and are commonly inserted as a stiffness potential[26,27]. We use a sinusoidal function, previously used in combination with the hydrophobic interaction described above[20]. This depends on the torsion angle between four consecutive residues, $\phi_i$ and has the form: $E_i^{stiff} = 0.5 \left(1 + \cos 3\phi_i\right) - 1$. This expression presents minima around $\phi_i = \pm 180°$ and $\phi_i = \pm 60$, so that extended conformations (similar to those found in $\beta$-sheets) and more compact ones (like those found in $\alpha$-helices) are reproduced. If the angle $\phi_i$ is too small (below 40°), the conformation is strongly penalized. See Ref.[20] for further details.

These contributions have been designed in terms of internal energetic units for simplicity. Note that having energy in internal units is very convenient in combination with Monte Carlo algorithms, as new configurations are accepted or rejected according to the detailed balance condition, that depends on the system temperature. Then, we do not have to use constants that would be cancelled out during our calculations. The used units are referred to a certain reference state of temperature $T_{ref}$ and energy $E_{ref} = k_B T_{ref}$ that can, if needed, be linked to a real experimental state.

Then, we have performed our simulations in terms of these reduced and adimensional units, defined as $T = T_{real}/T_{ref}$ and $E = E_{real}/k_B T_{ref}$. The balance among the contributions is defined in terms of the following weighting factors: $\omega^{hb} = 9.5$, $\omega^{stiff} = 7.0$ and $\omega^{hp} = 6.5$; it leads to the following global expression: $E = \omega^{hb} \, E^{hb} + \omega^{hp} \, E^{hp} + \omega^{stiff} \, E^{stiff}$.

### D. Effect of pH in peptide systems - Lessons from full atom simulations

Solution acidity is an important thermodynamic variable that can affect a biological system in a way as relevant as the effect of temperature or concentration. The system acidity is quantified through pH, that measures the concentration of protons (denoted by [H$^+$]) and is defined as pH= $-\log_{10}$[H$^+$]. Depending on this concentration, the protonation state of some amino acids of the protein may also change. Within a reasonable range, pH= [1, 13], backbone hydrogens (forming the peptide bond) are not affected[28] by changes in pH, but some side chains may change their protonation state (see Table III). This trend towards protonation is measured by the pK$_a$, pK$_a = -\log_{10}$K$_a$, where K$_a$ is the protonation constant of a certain acidic equilibrium as:

$$\text{AH} \overset{K_{a_\lambda}}{\rightleftharpoons} \text{A}^- + \text{H}^+$$

and can be defined as K$_a = \frac{[H^+][A^-]}{[AH]}$, where brackets indicate the concentration of each species. To evaluate the effect of protonation in peptides, we performed simulations at atomistic level (see Subsection II B) for two small test systems. The test systems consist of two peptides with different structures (an $\alpha$-helix, of PDB code 1RNU[29] and sequence KETAAAKFERQHM, and a $\beta$-hairpin, of PDB code 1E0Q[30] and sequence MQIFVKTLDGKTITLEV). We have chosen three extreme conditions: infinitely low pH (i.e. fully protonated lateral chains), physiological pH (usual protonation state of each lateral chain) and infinitely high pH (fully deprotonated chains).

Our intention, then, is to evaluate the impact of this variable on energetic and structural aspects of our systems, e.g. stability of native configurations, changes in energetic contributions, conformational differences. These analysis have been performed using facilities within the GROMACS package.

We show some results in Figure 3. In graphs (a) and (b) of this Figure we show the average and standard deviation of the electrostatic energy for the three pH conditions and our two peptides, computed from the coulombic term of the OPLS-AA force field. We can observe that the physiological pH exhibits a higher stability in both cases. Besides, the electrostatic interaction is strongly affected by this parameter, as its energy can change for more than 40 % (see Figure 3(a), from low to physiological pH values).

We have also evaluated the system hydrophobicity, as our coarse-grained model explicitly presents a hydrophobic interaction term. Since hydrophobicity is entropically driven, we have inferred the effect of pH in an indirect way. First, we have computed the energy associated to the Lennard-Jones term of the non-bonded interactions of the OPLS-AA force field, which measures the associated dispersive energy and can be partially ascribed to the system hydrophobicity. We can see in graphs (c) and (d) in Figure 3 that there is a slight decrease at non-physiological conditions. This energy difference is much smaller (less than 5%) than in the electrostatic energy, and comparable to the system fluctuations (related to the size of the error bars in the plots). A closer view to the stable structure in each case suggests that these small differences are linked to slight conformational changes to accommodate the charges, but not to the strengthening of each of them.

A further hint on the effect of pH in hydrophobicity can be obtained by the solvent acccesible surface area of the hydrophobic groups[31], shown in graphs (e) and (f) of Figure 3. We can see there that this value remains constant in all pH conditions; this result and the dispersive energy observations suggest that hydrophobicity is not strongly affected by pH.

Finally, we have evaluated the strength of a hydrogen bond in the three chosen conditions (graphs (g) and (h) in Figure 3). As we have already pointed out, it is generally accepted that the properties of backbone hydrogen bonds are not affected by the pH of the environment[28]. In these plots, the energy per hydrogen bond (calculated through the DSSP algorithm) is approximately constant in the two systems. Taking into account the definition of hydrogen bond in this algorithm, the result confirms that the hydrogen bond geometrical properties remain unchanged.

In a nutshell, we see that there is no significant difference in the hydrogen bond properties nor in the hydrophobic interaction stabilization due to the change in protonation state. However, as expected, we detect a major change in the electrostatic forces (see Figure 3(a) and Figure 3(b) and their scales). While the description of hydrogen bond and hydrophobic interactions is unaffected by the change in pH, a correct modelling of electrostatic interaction is instead required to build a valid pH-dependent model. In addition, the design of a pH-algorithm is essential to "tell" the system which pH conditions apply in a particular simulation.

### E. pH modelling

Modeling of the effects of the change in pH is a subject of intense research activity; currently different approaches coexist but most of them employ full-atom resolution[6,7] rather than coarse-grained models. For this reason our aim is to develop a coarse-grain strategy to tune the model presented in the previous sections in order to include the additional effects due to the change in pH. Therefore, we have chosen a discrete protonation state model (i.e. each site is either protonated or deprotonated) in which a Monte Carlo move may change the protonation state of a single site at each step (this idea is similar to that of Ref.[32]; instead for a review of alternative strategies, see Ref.[6]).

In our algorithm, each pH "move" selects a protonable site $i$ and changes its protonation state; next, it calculates the energy difference between the new and old configurations. The energy difference consists of two terms: One associated to the system interactions, $\Delta E_{int}$ (this depends on the way system's interactions are defined, in our specific case an explicit definition is given in the next Subsection), and one associated to the chemical energy, $\Delta E_{chem} = \ln(10)k_B T(\text{pH-p}K_a(i))$. Finally, the move is accepted or rejected according to the detailed balance condition, $P_{acc} = \min[1, \exp(-\Delta E/k_B T)]$, where:

$\Delta E = \Delta E_{chem} + \Delta E_{int} = \ln(10)k_B T[\text{pH-p}K_a(i)] + \Delta E_{int}$.

Note that, differently from the full-atom approach[32–35], we do not need an additional term for a reference free energy, as we are not explicitly including the proton, i.e. protonated and deprotonated beads are identical in terms of internal bonds; they differ only for the global charge. Because of this, we do not need preliminary thermodynamic integration steps or biasing functions[6]. $\text{p}K_a(i)$ is approximated with the experimental $\text{p}K_a$ of the isolated amino acid (see Table III); in fact we are working with relatively small peptides with a high solvent exposure, thus the simplification is rather reasonable and does not affect the overall behavior of our system.

We have tested the basic validity of our pH-dependent algorithm with numerical simulations of a non-interacting system, i.e. $\Delta E_{int} = 0$. This example is the simplest case we can consider, as protonation states are allowed to change depending only on their equilibrium constant (and not due to additional pH-dependent interactions). If the algorithm is reasonable, an average charge of the system according to the $\text{p}K_a$ values should be expected.

Our test system is, then, formed by one peptide of sequence: Ser-Glu-Val-Val-Lys-Ser-Val-Ser in implicit solvent. In this peptide, glutamic acid (Glu) maybe neutral or negative and lysine (Lys) may be positive or neutral, depending on the pH (see Table III). Therefore, in a system like this we could have a positive global charge (+1) if both glutamic acid and lysine are protonated, null (0) if glutamic acid and lysine have opposite charges or none of them was charged or negative (-1) if both pH-dependent amino acids are deprotonated.

We have evaluated how the global charge evolves as a function of the pH. We show in Figure 4(a) a comparison between the theoretically predicted charge, calculated from the equilibrium constant definition adapted to this particular example (see Subsection II D), and the average simulated charge (computed from our ensemble). We can see that our simulated values (blue dots) lie on the theoretical curve (black dotted line), proving that the behavior of the overall charge of our model matches the expected value.

In our model the average charge comes from an ensemble of discrete values; for each pH value, this distribution is plotted in Figure 4(b) by a colored bar. The different colors represent the population of each certain charge, normalized with respect to the whole equilibrium ensemble at a given pH.

The results of pH = 1, 7 and 13 show that 100% of the total population has a global charge +1, 0 or -1, respectively (we have also checked that the "uncharged" global state is formed by charged lysine and glutamic acid). This means that, at those pH values, the charge of each amino acid is essentially constant, which also justifies the connection between charge and pH that we used in our full-atom simulations (see Subsection II D). If pH is closer to the pK of either glutamic acid or lysine (marked with red vertical dots in the graph), we can see that at least two states coexist; we have observed that the amino acid that changes is always the one whose pK lies closer to the simulated pH value.

These numerical tests prove that the pH-dependent simulation algorithm is a reasonable starting point, as the shown results agree with theoretical data and reproduce the expected balance of protonation states of a non-interacting system.

## F. Electrostatic potential

The numerical test reported in the previous section confirms that the algorithm including the pH dependency is robust at the very basic level; next step, from the methodological point of view, is that of designing a coarse-grained model with pH-dependent interactions, $\Delta E_{int} \neq 0$. According to bibliographic data[28] and our findings in Subsection II D, that means modeling electrostatic interactions. We have chosen the Yukawa potential as a functional form of our potential because this is extensively used for electrostatic interactions in other contexts and because despite the simplicity of its functional form the screening effects can be efficiently described [36,37]. As for the case of atomistic electrostatic potentials, the specific form of the potential is pH-independent, i.e. the pH sensitivity comes from the change in the protonation state of the ensemble. The potential, $E^{elec}$, in then defined as follows:

$E^{elec} = A_0 * [\exp(-d/A_1)/d] + A_2$

$A_0$, $A_1$ and $A_2$ are adjustable parameters and $d$ is the distance between two given interacting beads. The key point in our modeling procedure is now the choice of the parameters.

Our initial guess is based on a classic strategy for potentials of mean force and knowledge-based strategies, that is, evaluating some properties of this interaction in real peptide systems[38]. We have taken a database of 1600 proteins[39] and we have searched for electrostatic interactions therein according to the definition in Ref.[40] (i.e. interacting atoms in lateral chains closer than 0.4 nm). For each of these pairs of amino acids, we have computed the distance $d$ between their $\alpha$-carbons (as this is the position of the interacting bead of our coarse-grained model).

We have built a histogram with these distances and applied to it the Boltzmann law, as shown in Figure 5a. This representation gives us a preliminary idea of which distances between electrostatically interacting amino acids are somehow favored in nature and, therefore, should also be favored in our potential.

We can observe in the plot that the minimum is a plateau below 1.0 nm. This means that, if there is an attractive electrostatic interaction in a protein, the distance between the involved $\alpha$-carbons will probably be smaller than this value. At higher distances, the values get gradually closer to zero, being negligible around 1.8 nm.

Using this preliminary information, we have optimized our Yukawa potential parameters by comparison with experimental data as well as full-atom simulations. Our parameters are: $A_0 = 22.6$, $A_1 = 15.0$ and $A_2 = 0.40$, resulting in the functional form plotted in Figure 5b. We have kept the plateau below 1.0 nm, using the Yukawa potential form between 1.0 nm and 1.8 nm. Our potential is truncated at that distance, avoiding in this way many complications derived from slow-decaying long range interactions[41]. Finally, we searched for a suitable weighting factor for our electrostatic interaction and obtained a value of $\omega^{elec} = 12.5$ in internal units.

The expression for the system interactions according to our global potential reads: $E = \omega^{hb} E^{hb} + \omega^{hp} E^{hp} + \omega^{stiff} E^{stiff} + \omega^{elec} E^{elec}$. Note that, in the case of a "pH-dependent move", only electrostatic interactions are modified. Therefore, $E_{int} = \omega^{elec} E^{elec}$.

The aforementioned parameters have been optimized by reproducing structural data from a set of peptides with bibliographic information available (see Section III). In general, for the modeling purpose, we have referred to different peptides and different conditions with the aim to design a potential that can be applied not only to the systems presented here, but also to other systems and conditions.

## III. RESULTS AND DISCUSSION

We have defined a coarse-grained potential that, together with a pH-dependent sampling algorithm, aims to describe the effect of pH in a peptide system. The validation of our model involves the simulation of a representative set of peptides and their comparison with bibliographic data (either experimental or simulated with full-atom resolution). We have selected several examples that illustrate different pH-related situations that can take place. The idea is to show that our specific model is characterized by a high degree of transferability (essentially due to the universality of the modeling strategy), i.e. that it can describe different peptides and external conditions. The choice of the examples has been mainly driven by the availability of detailed experimental data and the observation of different responses when changing the media acidity. We present three cases: in the first one, the peptide loses its native structure at non-physiological pH conditions; in the second one, its structure is better defined in an acidic environment; the third one presents a different aggregation propensity at low pH conditions.

Our first example is the C peptide from ribonuclease A (PDB code 1RNU and sequence KETAAAKFERGHM), a peptide with extensive pH-dependent information, both experimental and from full-atom simulations[42–44]. Having detailed data for the full range of pH values, we have carried out extensive single chain parallel tempering simulations at 13 different pH values ranging from 1 to 13. We have computed heat capacity curves in terms of temperature to evaluate the system stability and taken the transition temperature form the maximum of these curves, as shown on Table IV.

The key aspect of this peptide is the tight relationship between structure and pH. In fact, it is known that this peptide presents a partially helical structure at mild pH values (close to physiological) that drops when pH is disturbed[42,43], leading to a bell-shaped pH profile. Our results, shown in Figure 6, prove that our model also reproduce this behavior. In plot Figure 6(a) we have represented the helix content of the C-peptide at several pH values (at a temperature $T^*$ that is located slightly below the unfolding temperature, $T_m^*$, see Table IV for details); we can see that the helicity content is higher in the case of intermediate pH values (4-6) and it clearly drops beyond this range.

Using the helicity content, we have calculated the mean residue ellipticity at 222 nm, $[\theta_{222}]$. This value, that can also be measured experimentally by circular dichroism, has been computed using the method described in Refs.[42,45,46]. We have compared our results with bibliographic data in Figure 6(b). In this plot, note that both simulation data (our coarse-grained and also the full-atom results estimated from Ref.[42]) have been shifted by 2000 deg·cm²·dmol⁻¹, as performed in the original work in Ref.[42]. The purpose of this shift is to allow an easier comparison with experimental data. We have proved in Figure 6(c) that this shift is related to the temperature at which helicity properties have been measured (note that the REMC algorithm allows to compute a wide range of temperatures within the same simulation). We show in the latter plot the evolution of the helicity content along temperature at pH 5 for our simulation data and experimental results[43]; we observe that the two sets of data exhibit a similar dependency with temperature; therefore, we can conclude that, regardless the particular values, our results are comparable not only to those from fully detailed simulations, but also to the experimental ones.

Thanks to the detailed information available for this peptide, we can further investigate the source of the higher helicity at intermediate pH values. It is thought to be related to the formation of particular electrostatic interactions (called "salt bridges" in this context)[42,43]. In particular, one of the involved issues is the role of salt bridges between residues 1 (lysine) and 9 (glutamic acid) and between residues 2 (glutamic acid) and 10 (arginine) and how a decrease in the pH destabilizes the peptide structure.

Although original works suggest the importance of the latter[43], more recent investigations highlight the pH sensitivity of the former one[42,44]. To investigate this aspect, we have estimated the free energy of the peptide at intermediate and low pH values (ranging 1-7) as a function of the distance between the involved residues by means of the Weighted Histogram Analysis Method (WHAM)[47].

We can see in Figure 7(a) the evolution in terms of distance Lys1-Glu9, that is clearly sensitive towards pH. In these plots, a smaller free energy value indicates a higher stability. We can observe that its minimum at low pH values (blue colors) is found at larger distances, implying that the most favorable configurations just present a weak or inexistent interaction between these residues. At intermediate pH values (red), however, the minimum lies around 0.86 nm; this distance correspond to a maximum strength in the electrostatic interaction of our potential, so our results indicate a key role of the interaction between residues 1 and 9 at intermediate pH values, in agreement with Refs.[42,44]. In the case of distance Glu2-Arg10, shown in Figure 7(b), its minimum is always kept at the same position, which implies a less active role in pH changes.

These evidences altogether show that our combination of a pH-dependent algorithm and a coarse-grained potential are able to describe in realistic terms the behavior of this pH-dependent peptide in accordance with experimental information, also comparable to full-atom simulations. The next step, then, is to test other examples to prove the methodology in different situations. The second example we show is hemagglutinin (particularly, we use the so-called loop-36, of sequence RVIEKTNEKFHQIEKEFSEVEGRIQDLEKYVEDTKI, that can be found in PDB code 1RUY), a peptide whose biological function is mediated by a conformational change that depends on pH[48].

In the case of this peptide the inverse pH-response is observed: it mainly forms a random coil at physiological conditions, but presents a stable helical structure at low pH values[49]. In this case we have simulated our peptide in infinite dilution conditions for two pH values (also experimentally measured in Ref.[49]): 4.5 (where a helical conformation is claimed to be more stable) and 7.2 (close to physiological pH, mainly unstructured).

We have found in our simulations that the helical content of hemagglutinin at low pH is around $0.52 \pm 0.01$, while this values drops down to $0.41 \pm 0.01$ at physiological conditions. In this way, the helicity of the peptide increases at low pH values, as experiments also report[48,49]. This shows that our methodology does not introduce an unnatural bias in our data (as it would happen, for instance, is low pH always lead to denaturated structures) but is sensitive to the particular features of the studied peptide. According to the kind of interactions in our model, we may state that this response is mediated by the sequence of the peptide itself, i.e. a high proportion of negatively charged residues (10 out of 36 total residues) at physiological pH, that become neutral at low pH conditions and, therefore, are no longer able to form stabilizing electrostatic interactions with the positively charged residues in the sequence. As a result, a higher helicity has been found in experiments and our simulations at low pH conditions, where this stabilization can take place.

So far, we have presented a system with thorough experimental and atomistic detailed information where non-physiological pH environments lower the helical native structure of the peptide (C peptide). We have also presented another system that, due to sequence effects, is stabilized at high pH values (hemagglutinin). Finally, we would also like to show an example that highlights the link between pH and aggregation. It is an octapeptide of the human prion protein (PDB code 1EOI of sequence HGGGWGQP)[50], involved in so-called folding diseases and known to form aggregates in high concentration and low pH conditions[50,51]. In this case, multichain simulations have been carried out at two different pH values (4.5 and 7.8, in order to compare our results with those in Ref.[50]. We have used relatively high concentration conditions (10 g/L aprox. in our simulation box) to favor the formation of multichain structures.

As for the rest of our simulations, we have taken the transition temperatures of each system to carry out a detailed comparison between the two pH conditions (4.5 and 7.8).We have observed that the number of interchain interactions is systematically higher at low pH conditions. This feature has allowed us to carry out a clustering procedure in terms of structural and energetic properties such as radius of gyration, root mean square deviation, distance among interacting beads, type of interactions among beads or chains, etc.

We have identified four different situations among our equilibrium configurations: (a) groups of peptides (formed by 2-5 independent chains) stabilized by interchain hydrogen bonds (also called "$\beta$-type aggregates", as they presented a very characteristic arrangement similar to a $\beta$-sheet), (b) groups of peptides (having 2-7 chains together) mainly stabilized by hydrophobic and electrostatic interactions ("disordered aggregates") and (c) isolated peptides (mainly helical), as well as denatured configurations (where just sporadic interactions are found and random configurations are predominant). A characteristic example of these clusters can be found in Figure 8, where each chain has been colored in a different way for the sake of clarity. Interestingly, structures similar to (a) and (b) have been found in experimental aggregation studies[52]. Structures (a) and (b) are oligomeric complexes (or multichain bodies) that do not present native conformation and, therefore, can be considered "aggregates"[53]. In structure (a) we observe a certain order, mediated by the formation of interchain hydrogen bonds, that provide a $\beta$-sheet shape. This kind of structured aggregates can be related to amyloids because of this sheet-like arrangements. Structure (b) presents an amorphous shape, due to the presence of many hydrophobic interactions; these collapsed structures are often related to inclusion bodies and other amorphous and insoluble protein formations.

At low pH (4.5) we have found that most of our configurations (75%) exhibit interactions among chains (40% belong to the $\beta$-type one and 35% are disordered aggregates, corresponding to situations (a) and (b), respectively). However, at physiological pH values (7.8) this percentage drops down to 30% (15% $\beta$-type and 15% disordered), showing how pH conditions induce the system formation of multipeptide structures. Interestingly, a connection between the size of the observed species and pH has also been reported in hydrodynamic experiments of this peptide[50]. An increase in the apparent molecular mass (i.e. formation of multi-peptide arrangements) at low pH values has also been reported in the full prion protein, where the authors link this behavior to an active role of the octapeptide we are studying[54]. This fact is common to other prion peptides and prion proteins[51]; furthermore, the relationship between low pH values and a higher aggregation propensity seems to be a protein hallmark[55]. The results of this latter study shows that also for systems where interchain structures is the main feature, our model can reproduce its essential feature with satisfactory consistency with respect to data from literature.

## IV. CONCLUSION

Employing a modeling procedure based on the consistency across the different scales (or molecular resolutions) we have developed a coarse-grained model of peptides where the effects of the changes in pH can be efficiently described. Although some approaches to the problem can be found at full-atom scale, this is the first time -to our knowledge- that a pH-dependent coarse-grained model has been described. We have tested it in a rather broad variety of situations and have shown that our results comply in general terms with atomistic simulations and experimental data available. Our choice of parameters has been kept constant throughout all the results present here, suggesting the universality of this methodology to simulate other peptides and pH conditions.

In addition, the performance of extensive multichain simulations (like in the case of the prion peptide) suggests the suitability of this strategy to carry out larger and more complex numerical experiments on these and other cases. In fact, as a future perspective, we will employ such a model to study aggregations in systems consisting of large peptides and study in particular the initial nucleation step as a function of the pH; this kind of study is currently not affordable with atomistic and/or other coarse-grained models available. Given the capability of predicting in a rather reasonable way, intra- and inter-chain structural properties of this model as a function of the pH, one may be confident that some essential feature of the nucleation process can be captured, and in turn eventually used as pilot information for experimental investigations.

**Tables**

TABLE I. Optimal ranges for the three geometrical restrictions chosen in our model for backbone hydrogen bonds.

| Restriction | Local range | Non-local range |
|---|---|---|
| R1 | 4.7 Å ≤ R1 ≤ 5.6 Å | 4.0 Å ≤ R1 ≤ 5.6 Å |
| R2 | 0.74 ≤ R2 ≤ 0.93 | 0.75 ≤ R2 ≤ 1.00 |
| R3 | 0.92 ≤ R3 ≤ 1.00 | 0.94 ≤ R3 ≤ 1.00 |

TABLE II. Values of the parameters of the sequence-dependent terms of $E^{hp}$. The right hand side contains the values that $\sigma_{int}$, $S_1$ and $S_2$ take depending of the kind of non-local interaction (note that the two latter quantities are adimensional).

| Type of interaction | $\sigma_{int}$/nm | $S_1$ | $S_2$ |
|---|---|---|---|
| Hydrophobic-hydrophobic | 0.40 | 15.45 | 1 |
| Polar-polar or hydrophobic-polar | 0.30 | 15.45 | -1 |
| Others | 0.30 | 15.45 | 0 |

TABLE III. pKa values of the side chains of pH-sensitive residues

| Positive amino acids | | Negative amino acids | |
|---|---|---|---|
| Name | p$K_a$ | Name | p$K_a$ |
| Arginine | 12.48 | Aspartic acid | 3.65 |
| Cysteine | 8.18 | Glutamic acid | 4.25 |
| Lysine | 10.53 | Histidine | 6.00 |
| Tyrosine | 10.07 | | |

TABLE IV. Temperature data for the C peptide, in reduced units. Transition temperature values, $T_m^*$, taken from the maximum in the heat capacity curve in our simulations. Temperature values, $T^*$, at which helicity has been measured.

| pH | $T_m^*$ | $T^*$ |
|----|---------|-------|
| 1  | 2.34    | 2.32  |
| 2  | 2.34    | 2.32  |
| 3  | 2.34    | 2.32  |
| 4  | 2.34    | 2.32  |
| 5  | 2.34    | 2.32  |
| 6  | 2.38    | 2.34  |
| 7  | 2.38    | 2.34  |
| 8  | 2.38    | 2.34  |
| 9  | 2.38    | 2.34  |
| 10 | 2.40    | 2.38  |
| 11 | 2.40    | 2.38  |
| 12 | 2.40    | 2.38  |
| 13 | 2.40    | 2.38  |

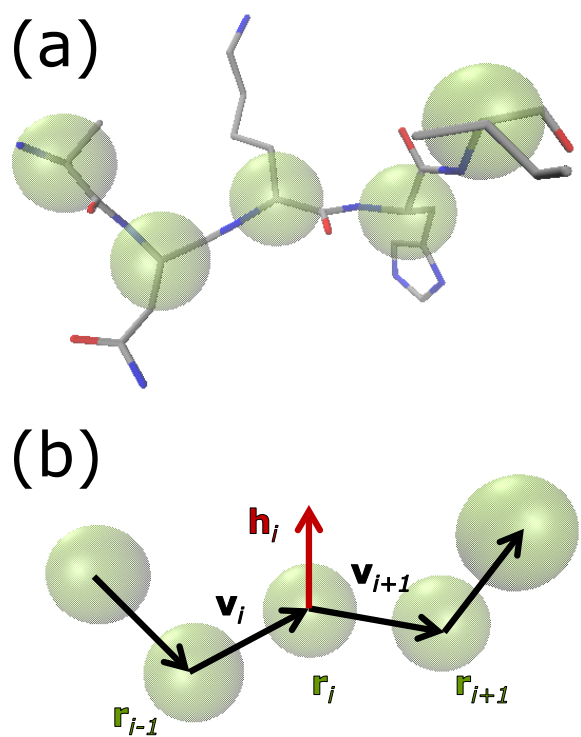

FIG. 1. Simplified representation used in our coarse-grained model. (a) Peptide represented with full-atom resolution with coarse-grained interacting beads (in green) superimposed. (b) Vectorial definitions needed for the coarse-grained potentials.

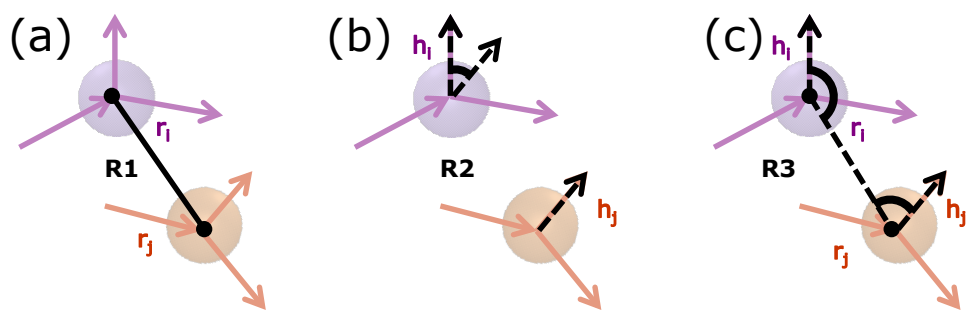

FIG. 2. Representation of the restrictions of the hydrogen bond interaction potential. (a) R1. (b) R2, (c) R3.

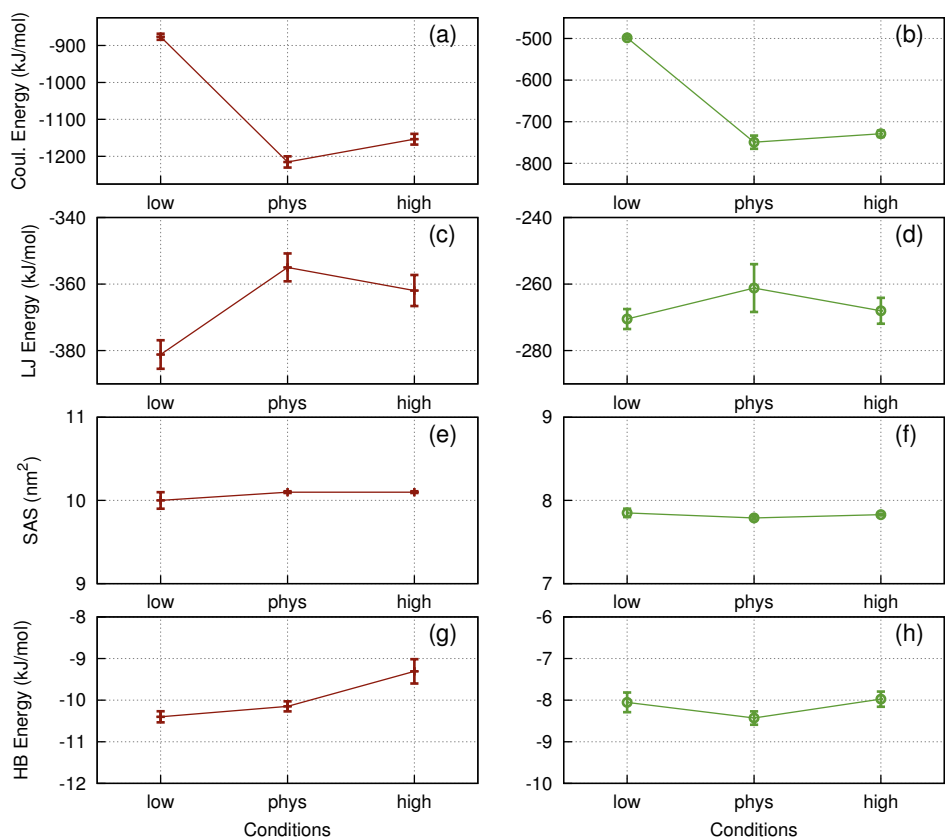

FIG. 3. Full-atom simulation data for two model peptides (1E0Q and 1RNU, in red and green, respectively) in three different protonation situations: low pH (fully protonated), physiological pH (usual protonation state found in physiological conditions) and high pH (fully deprotonated). (a) and (b) Coulombic energy. (c) and (d) Dispersive Lennard-Jones energy. (e) and (f) Solvent accesible surface area. (g) and (h) Energy per hydrogen bond.

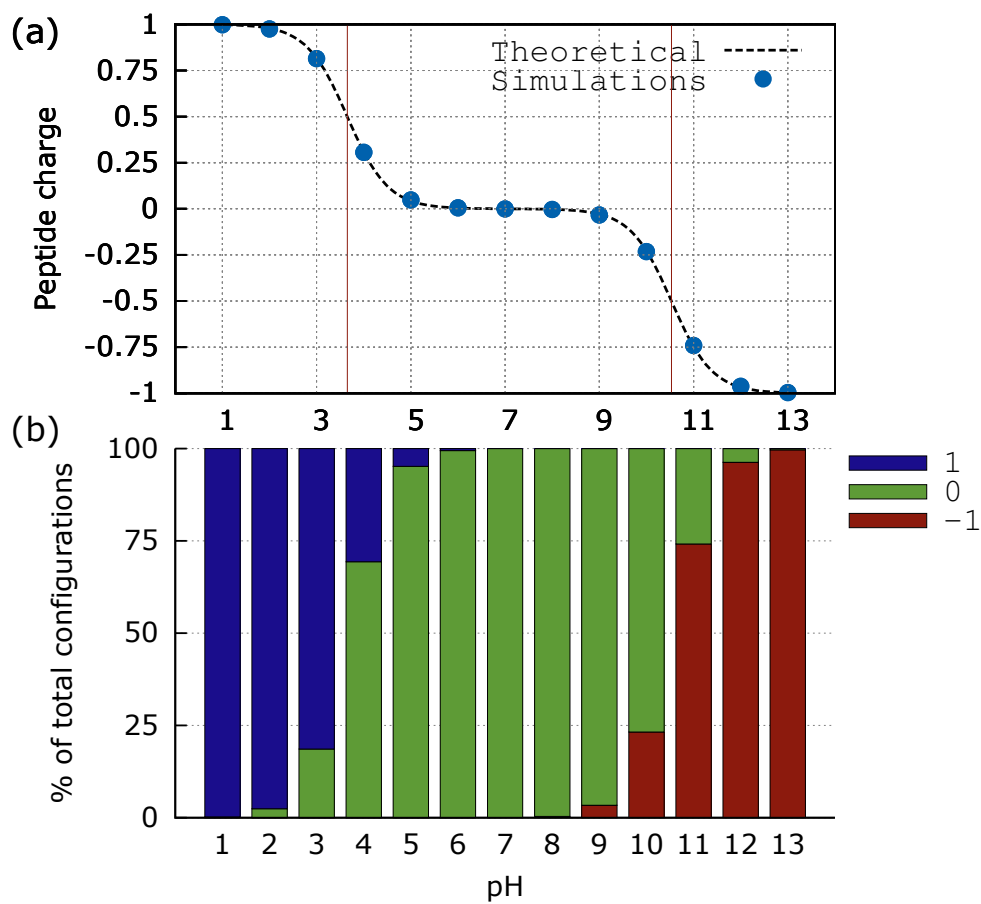

FIG. 4. Numerical tests on our constant-pH simulations algorithm using a non interacting peptide (see text for details). (a) Theoretical and simulated total charge of our system as a function of pH. For the sake of clarity, we have marked with red vertical lines the $pK$ values of the pH-sensitive residues of the peptide. (b) Evolution of the system charges as a function of pH for the system ensemble.

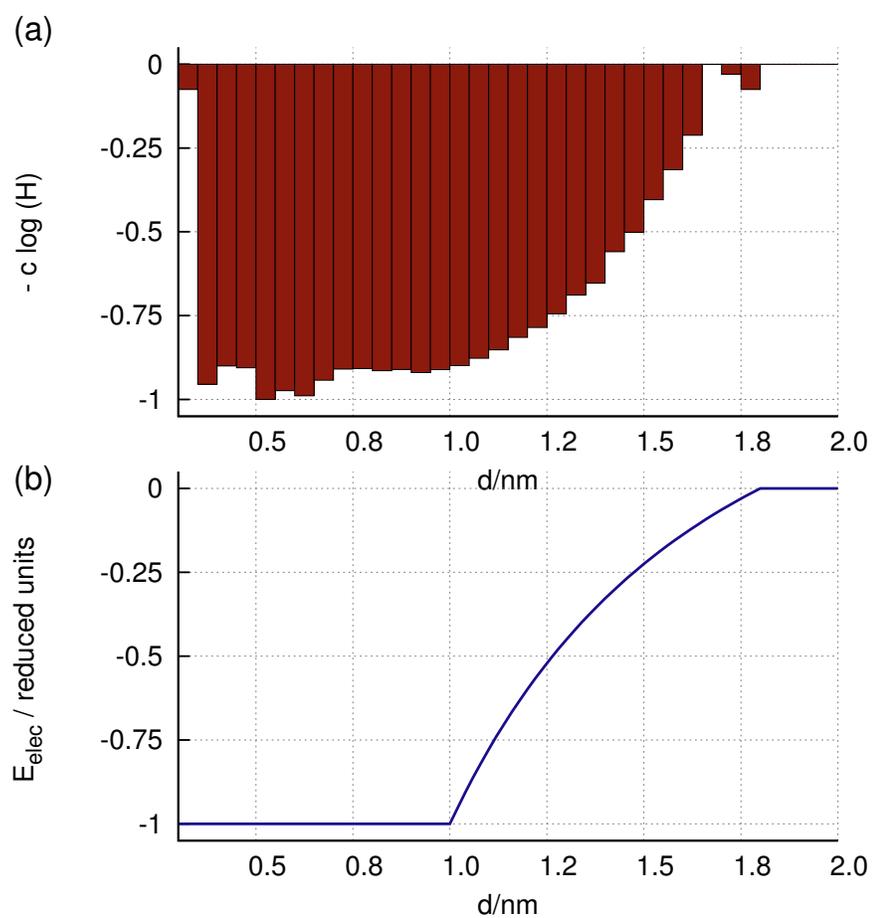

FIG. 5. Information about electrostatic potential. (a) Boltzmann law on experimental structures for pairs interacting through electrostatic forces (see text for further details). (b) Representation of the functional form of our electrostatic potential, following the Yukawa expression.

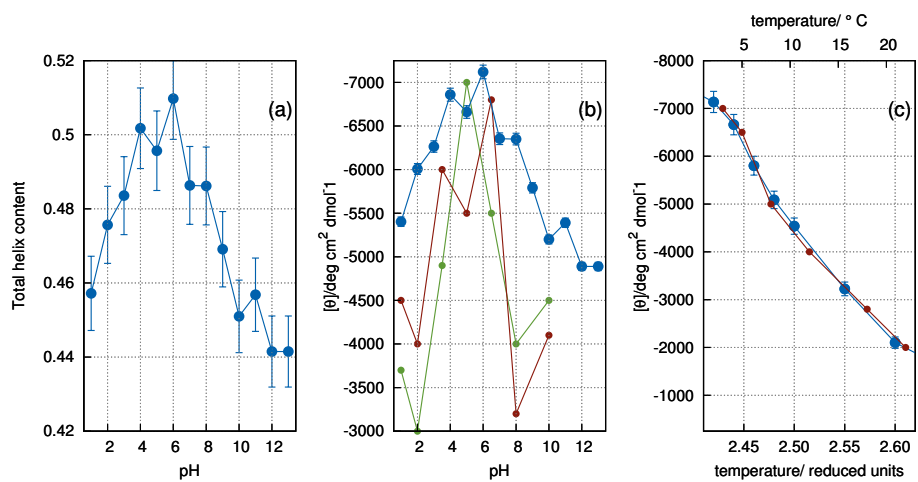

FIG. 6. Relationship between helicity and pH for the C peptide. (a) Total helix content of our coarse-grained simulations. (b) Ellipticity of coarse-grained simulations (blue), full-atom simulations from Ref.[42] (red) and experimental data from Ref.[43] (green). (c) Evolution of ellipticity at pH 5 with temperature in our simulations (blue, in terms of reduced temperature units) and in circular dichroism experiments (red, temperature in degrees centigrades)[43].

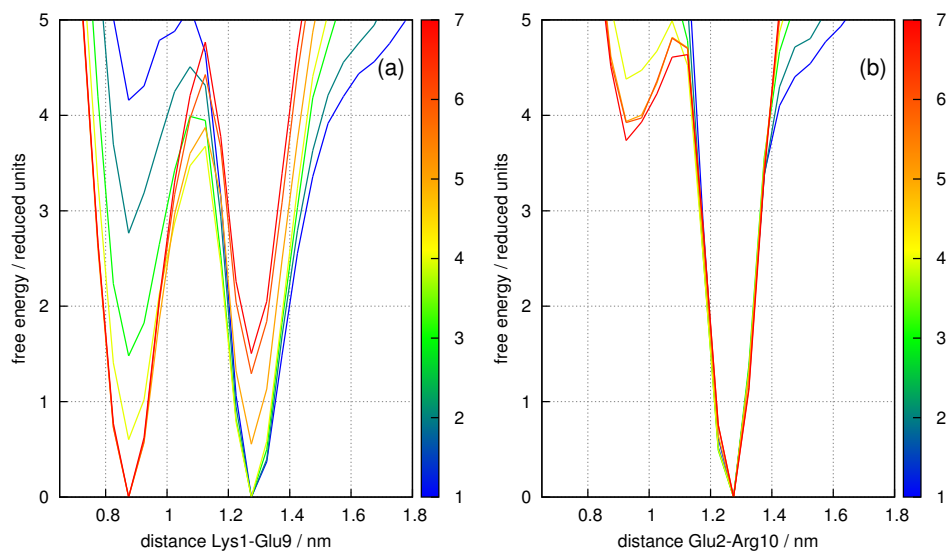

FIG. 7. Free energy profiles of the C peptide at pH values ranging 1 (blue) to 7 (red) in terms of (a) distance between Lys1 and Glu9 and (b) Glu2 and Arg10.

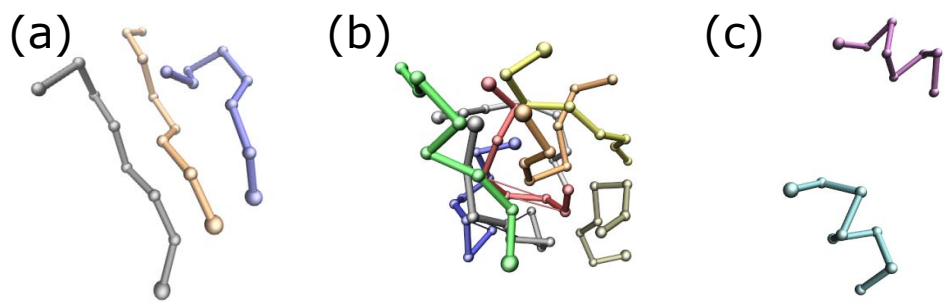

FIG. 8. Representative structures found in the prion peptide multichain simulations. (a) $\beta$-type cluster (note the parallel arrangement among chains, hallmark of interchain hydrogen bonds). (b) Disordered cluster (chains glued together by means of electrostatic and hydrophobic interactions that have spherical symmetry and, thus, favor disordered round-like arrangements. (c) Isolated chains that form $\alpha$-helices.